\documentclass[prl,aps,letterpaper,floatfix,superscriptaddress,preprintnumbers,twocolumn,10pt]{revtex4-1}

\usepackage{dcolumn}
\usepackage{bm}
\usepackage[dvips]{graphicx}
\usepackage{epsfig}
\usepackage{amsmath, amsfonts, amssymb}

\usepackage{multirow}
\usepackage{hyperref}
\usepackage{pifont}

\newcommand{\be}{\begin{equation}}
\newcommand{\ee}{\end{equation}}
\newcommand{\bee}{\begin{equation*}}
\newcommand{\eee}{\end{equation*}}
\newcommand{\bea}{\begin{eqnarray}}
\newcommand{\eea}{\end{eqnarray}}
\newcommand{\bean}{\begin{eqnarray*}}
\newcommand{\eean}{\end{eqnarray*}}

\usepackage{xcolor}
\definecolor{red}{rgb}{1,0,0}
\definecolor{purple}{rgb}{0.5,0,0.5}
\definecolor{blue}{rgb}{0,0,1}
\definecolor{green}{rgb}{0,1,0}

\begin{document}

\title{Probing the Origin of Neutrino Masses and Mixings via Doubly Charged Scalars:\\
Complementarity of the Intensity and the Energy Frontiers}

\author{Tanja Geib}
\affiliation{Max-Planck-Institut f\"ur Physik (Werner-Heisenberg-Institut), 
F\"ohringer Ring 6, 80805 M\"unchen, Germany}
\author{Stephen~F.~King}
\affiliation{Physics and Astronomy, University of Southampton, Southampton, SO17 
1BJ, United Kingdom}
\author{Alexander~Merle}
\affiliation{Max-Planck-Institut f\"ur Physik (Werner-Heisenberg-Institut), 
F\"ohringer Ring 6, 80805 M\"unchen, Germany}
\affiliation{Physics and Astronomy, University of Southampton, Southampton, SO17 
1BJ, United Kingdom}
\author{Jose~Miguel~No}
\affiliation{Department of Physics and Astronomy, University of Sussex, Brighton 
BN1 9QH, United Kingdom}
\author{Luca~Panizzi}
\affiliation{Physics and Astronomy, University of Southampton, Southampton, SO17 
1BJ, United Kingdom}

\date{\today}

\begin{abstract}
We discuss how the intensity and the energy frontiers provide complementary constraints within a minimal model of neutrino 
mass involving just one new field beyond the Standard Model at accessible energy, namely a doubly charged scalar $S^{++}$ and its 
antiparticle $S^{--}$. In particular we focus on the complementarity between high-energy LHC searches and low-energy probes such as lepton flavor violation. 
Our setting is a prime example of how high- and low-energy physics can cross-fertilize each other.
\end{abstract}

\preprint{MPP--2015--299}
\maketitle


The origin of neutrino mass and mixing is an outstanding open question, as the existence of massive neutrinos, which follows from the discovery of 
neutrino oscillations~\cite{Fukuda:1998mi,Ahmad:2002jz}, cannot be accommodated within the Standard Model (SM) of particle physics. Together with other puzzles 
like the nature of Dark Matter or the generation of the observed matter-antimatter asymmetry in the Universe, this constitutes a leading motivation to 
search for new physics beyond the SM. 

The experimental effort to unravel the nature and properties of such new physics is pursued along three main avenues: the \emph{energy}, \emph{intensity}, 
and \emph{cosmic} frontiers, which provide highly complementary probes of new physics. A prime example of such a complementarity arises if the new physics 
responsible for neutrino masses and mixings lies not very far above the electroweak scale, case in which both low and high-energy experiments could be 
sensitive to it. However, even though there are well-motivated scenarios for the generation of neutrino masses and mixings which predict signatures at both 
the intensity and high-energy frontiers, such as low-scale seesaw models (see {\it e.g.}\ the discussion in \cite{Mohapatra:2005wg}) and loop-induced 
models~\cite{Zee:1980ai,Babu:1988ki,Krauss:2002px,Ma:2006km,Aoki:2008av,Gustafsson:2012vj}, these scenarios generically predict the existence of 
multiple new particles, making concrete predictions for phenomenology difficult to extract. 

A doubly charged scalar particle $S^{++}$ is predicted in a large class of these scenarios 
in connection to Lepton Number Violation (LNV) and the generation of neutrino masses. An extension of the SM by \emph{just one new particle} at 
accessible energy, $S^{++}$ (being $SU(2)_L$ singlet to avoid the introduction of extra degrees of freedom from an $SU(2)_L$ multiplet), and in 
the presence of effective operators giving rise to LNV, provides \emph{the most minimal framework} which captures the main features of a large 
class of neutrino mass models~\cite{King:2014uha}. It allows to fully exploit the complementarity between collider searches and low-energy probes such 
as lepton flavor violating (LFV) processes.

In this work we manifestly explore the complementarity of the two experimental avenues as a probe of a doubly charged (but $SU(2)_L$ singlet) 
scalar particle $S^{++}$. From the low-energy perspective (\emph{intensity} frontier), while the experimental limits on the LFV 
processes $\mu \to e\, \gamma$ from MEG~\cite{Adam:2013mnn} and $\mu \to 3\, e$ from SINDRUM~\cite{Bellgardt:1987du} are at present 
the most stringent ones, the most dramatic upcoming experimental advances are to occur in $\mu^-$--$e^-$ conversion in nuclei, which 
is expected to become the most sensitive LFV probe in the future~\cite{Raidal:2008jk}, 
expected to reach a sensitivity to branching ratios of $10^{-17}$ already in the nearer future~\cite{Cui:2009zz}. 
At the same time, the Large Hadron Collider (LHC) will probe during 
Run-II the TeV region for new particles like $S^{++}$. It is these two probes that we concentrate on here, assessing their respective 
reach in these scenarios. Our discussion reveals how strongly information from both the \emph{intensity} and \emph{energy} frontiers can 
complement each other, to maximize our benefit from on-going and near-future experiments searching for new physics. 

\vspace{-2mm}

\section{Intensity/Energy Complementarity:\\ A Key Approach to Neutrino Masses}
\vspace{-2mm}

\begin{figure}
\centering
\includegraphics[width=7.cm]{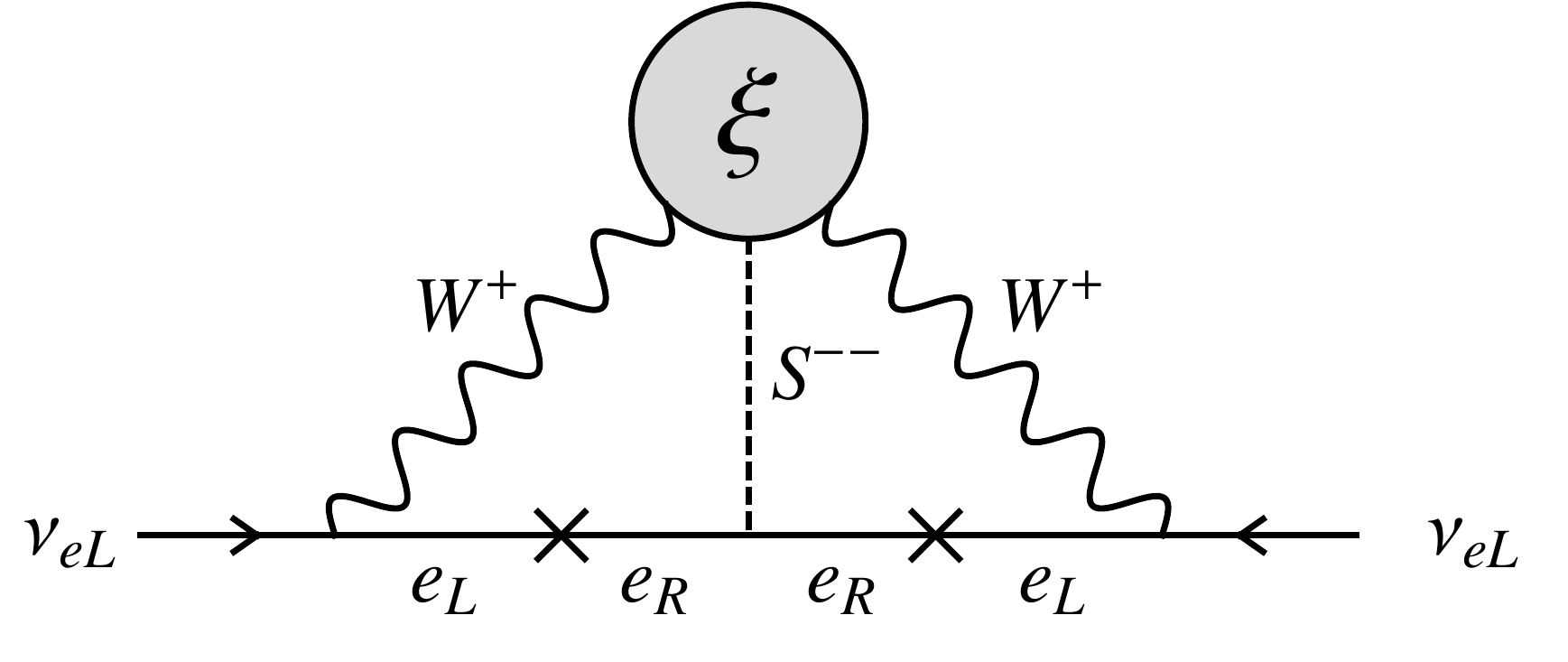}
\caption{\label{fig:NeutrinoMass}Neutrino mass generation for the example of the $m_{ee}^\nu$ element of the full neutrino mass matrix. 
Similar diagrams are responsible for the full matrix $m^\nu_{ab}$, where $a,b=e,\mu,\tau$.}
\end{figure}

Let us start by discussing the theoretical framework for the SM with the addition of an $SU(2)_L$ singlet, doubly charged 
scalar $S\equiv S^{++}{\rm or}\ S^{--}$. Our renormalizable Lagrangian is
\be
\label{Lagrangian}
\mathcal{L} = \mathcal{L}_{\mathrm{SM}} + (D_{\mu}S^{++})^{\dagger}(D^{\mu}S^{++}) + f_{ab}\, \overline{(\ell_{R})_a^c} \ell_{Rb}\, S^{++}
+ \mathrm{h.c.},
\ee
with $a,b = e,\mu,\tau$ being flavor indices and $f_{ab}$ a symmetric matrix in flavor space. The scalar $S^{++}$ is assumed to 
have a mass $M_S$. The Lagrangian \eqref{Lagrangian} conserves lepton number, thus \emph{not} leading to neutrino mass generation. 
However, allowing for non-renormalizable operators containing both $S^{++}$ and SM fields leads to LNV. Assuming that $S^{++}$ is 
connected to the generation of neutrino masses (which by construction forbids the $D = 5$ Weinberg operator), the leading LNV operator 
appears at $D=7$~\cite{King:2014uha} (see also~\cite{delAguila:2012nu,Gustafsson:2014vpa}):
\be
\label{SppEffLNVO9}
\frac{\xi}{\Lambda^3}\,\left[H^T i\sigma_2 \left(D_{\mu} 
H\right)\right]\,\left[H^T i\sigma_2 \left(D^{\mu} H\right)\right]\, S^{++} + 
\mathrm{h.c.},
\ee
which leads to an interaction $S^{\pm\pm}\,W^{\mp}W^{\mp}$. Combined with the last term in~\eqref{Lagrangian}, this breaks lepton 
number by two units and at two-loop order gives rise to light neutrino masses of Majorana nature, as shown in Figure~\ref{fig:NeutrinoMass}, 
by adding \emph{only one new particle} to the SM.

We stress that the interactions in (\ref{Lagrangian}) suffice to describe the physical processes which allow to fully exploit the 
low-high energy complementarity in this class of scenarios. Nevertheless, the key for complementarity is that the matrix $f_{ab}$ is 
far from arbitrary, as it enters into the generation of the neutrino mass matrix $(m^\nu_{ab})$: 
\begin{equation}
\label{eq:nu-mass}
(m^\nu_{ab}) \sim
\begin{pmatrix}
m_e^2 f_{ee} & m_e m_\mu f_{e\mu} & m_e m_\tau f_{e\tau}\\
m_e m_\mu f_{e\mu} & m_\mu^2 f_{\mu\mu} & m_\mu m_\tau f_{\mu\tau}   \\
m_e m_\tau f_{e\tau} & m_\mu m_\tau f_{\mu\tau} & m_\tau^2 f_{\tau\tau} 
\end{pmatrix},
\end{equation}
with the symmetric nature of $(m^\nu_{ab})$ being just a reflection of the light neutrinos being Majorana particles. 
The structure of $(m^\nu_{ab})$ is constrained by the measurements of all light neutrino mass squared differences and 
leptonic mixing angles by neutrino oscillation experiments~\cite{Gonzalez-Garcia:2014bfa,Bergstrom:2015rba}. Combining these 
with current bounds on $f_{ab}$ and $M_S$ from LFV processes like $\mu \to e\gamma$~\cite{Lee:1977tib,Cheng:1980tp,Lavoura:2003xp} and the 
LNV neutrinoless double $\beta$-decay process, we extract three representative average sets of couplings $\{f_{ab}\}$~\cite{King:2014uha}, 
called \textcolor{red}{\bf red} ($f_{ee} \simeq f_{e\tau} \simeq 0$), 
\textcolor{purple}{\bf purple} ($f_{ee} \simeq 0$ \& $|f_{e\mu} f_{\mu\mu}^*| \simeq |f_{\mu\tau}^* f_{e\tau}|$), 
and \textcolor{blue}{\bf blue} (only $|f_{e\mu} f_{\mu\mu}^*| \simeq |f_{\mu\tau}^* f_{e\tau}|$) in Tab.~\ref{Tab:Scenarios}.

\begin{table}[h!]
\centering
\renewcommand{\baselinestretch}{1.25}\normalsize
 \begin{tabular}{c|c|c|c|c|c|c}
  & $f_{ee}$ & $f_{e\mu}$  & $f_{e\tau}$ &  $f_{\mu\mu}$ & $f_{\mu\tau}$ &  $f_{\tau\tau}$ \\
  \hline
 \textcolor{red}{\bf red} & $< 10^{-10}$  &  $10^{-2}$ & $< 10^{-10}$ &   $10^{-4}$ & $10^{-5}$ & $10^{-6}$ \\
 \hline
 \textcolor{purple}{\bf purple} & $< 10^{-10}$ &  $10^{-3}$  & $10^{-2}$ & $10^{-3}$ & $10^{-4}$ & $10^{-5}$ \\
\hline
\textcolor{blue}{\bf blue}  & $10^{-1}$   &  $10^{-4}$  & $10^{-2}$ & $10^{-3}$ & $10^{-4}$ & $10^{-5}$ \\
\hline
 \end{tabular}
 \renewcommand{\baselinestretch}{1.00}\normalsize
 \caption{\label{Tab:Scenarios}Benchmark sets of couplings for the possible neutrino mass scenarios.}
\end{table}


This is the ideal situation from the point of view of complementarity between LHC searches and low-energy LFV probes, from 
which $\mu^-$--$e^-$ conversion in nuclei is going to be the most powerful in the upcoming future. For each set 
of couplings (\textcolor{red}{\bf red}, \textcolor{purple}{\bf purple} and \textcolor{blue}{\bf blue}) the only 
free parameter in both cases is the mass of the new particle $S^{++}$. In the following we discuss the analyses for 
both, and the resulting complementarity, for both the benchmark slopes 
(displayed as \textcolor{red}{\bf red}, \textcolor{purple}{\bf purple} and \textcolor{blue}{\bf blue} lines corresponding 
to the representative sets of couplings in Tab.~\ref{Tab:Scenarios}) and for benchmark points (displayed 
as \textcolor{red}{\bf red}, \textcolor{purple}{\bf purple}, and \textcolor{blue}{\bf blue} dots) taken from~\cite{King:2014uha}, 
where all slopes and points correctly describe the neutrino masses and mixings.

\vspace{-2mm}

\subsection{Intensity Frontier: $\boldsymbol{\mu^-}$--$\boldsymbol{e^-}$ Conversion in Nuclei}

\vspace{-2mm}

We now explore the process of $\mu^-$--$e^-$ conversion on a nucleus~\cite{Weinberg:1959zz,Marciano:1977cj} in the 
presence of $S^{++}$ and its antiparticle $S^{--}$. We discuss here the main results and present the technical details 
of the calculation elsewhere~\cite{Proj:LFV}. The branching fraction of $\mu-e$ conversion with respect to ordinary muon 
capture rate $\Gamma_{\mathrm{Capt}}$, in the limit that the long-range contributions (mediated by the photon $\gamma$ in Figure~\ref{fig:LFV}) 
dominate the process, can be written as:
\begin{equation}
\mathrm{BR(}\mu^- N \to e^- N\mathrm{)} \simeq \frac{8\,\alpha_{\mathrm{EM}}^5 \,m_{\mu}\,Z^4_{\mathrm{eff}}\,F^2_p}{\Gamma_{\mathrm{Capt}}}\,\Xi^2_{\mathrm{particle}},
\end{equation}
with the effective atomic charge $Z_{\mathrm{eff}}$ and the nuclear matrix element $F_p$. The term $\Xi_{\rm particle}$ encodes the particle physics part 
of the amplitude. The above factorization of the branching fraction into particle and nuclear physics parts is made possible precisely by neglecting 
the short-range (non-photonic) contributions to the amplitude (Figure~\ref{fig:LFV}, Right), which are found to be parametrically suppressed by 
roughly $\mathcal{O}(m_{\mu}^2/M_W^2) \sim 10^{-5}$~\cite{Proj:LFV}. This factorization is very convenient, since all nuclear physics uncertainties 
and isotopic dependences can be absorbed into the experimental bounds. The particle physics amplitude $\Xi_{\rm particle}$ is given by~\cite{Proj:LFV}:
\begin{eqnarray}
 && \Xi_{\rm particle} = \frac{| \Pi_e f_{ee}^* f_{e\mu} + \Pi_\mu f_{e\mu}^* f_{\mu\mu} + \Pi_\tau f_{e\tau}^* f_{\tau\mu} |}{12\sqrt{2} \pi^2 m_\mu M_S^2}, 
 \label{eq:muecon_particle-part}\\
 && {\rm with}\ \ \Pi_a = 4 m_a^2 m_\mu - m_\mu^3 [ 1 -\ln (m_a^2/M_S^2)] \nonumber\\
 && + 2 (m_\mu^2 - 2 m_a^2) \sqrt{m_\mu^2 + 4 m_a^2}\ {\rm Arctanh} [m_\mu/\sqrt{m_\mu^2 + 4 m_a^2}].\nonumber
\end{eqnarray}
The structure of $\mu$--$e$--$\gamma$ in the middle diagram of Figure~\ref{fig:LFV} enters the loop-functions $\Pi_{e,\mu,\tau}$. 

While naively one could expect not much to change compared to $\mu \to e\gamma$, whose amplitude is proportional 
to $C\cdot |f_{ee}^* f_{e\mu}+f_{e\mu}^* f_{\mu\mu}+f_{e\tau}^* f_{\tau\mu}|/M_S^2$ (where $C$ is a constant incorporating 
all numerical factors), the coefficients $\Pi_a$ ($a = e,\mu,\tau$) in~\eqref{eq:muecon_particle-part} cannot be factored out. 
This immediately explains why the corresponding bound will be very strong: the benchmark scenarios described in~\cite{King:2014uha} 
all avoid the bound from $\mu \to e\gamma$ by relying on some cancellation among the couplings in the expression 
$(f_{ee}^* f_{e\mu}+f_{e\mu}^* f_{\mu\mu}+f_{e\tau}^* f_{\tau\mu})$. This cancellation is spoiled if there is no 
common prefactor $C$ anymore. Thus, when taking into account the experimental improvements in searches for $\mu^-$--$e^-$ conversion, 
this process provides a very strong bound on the otherwise perfectly working scenarios.

The resulting bounds on the scenarios found in~\cite{King:2014uha} are displayed in Figure~\ref{fig:Scenarios}. The model 
predictions are illustrated in two ways, for actual benchmark points 
(displayed as \textcolor{red}{\bf red}, \textcolor{purple}{\bf purple}, and \textcolor{blue}{\bf blue} dots) taken from~\cite{King:2014uha}, and for 
the representative sets of couplings in Tab.~\ref{Tab:Scenarios} (displayed as \textcolor{red}{\bf red}, 
\textcolor{purple}{\bf purple}, and \textcolor{blue}{\bf blue} lines), which comprise ``averaged'' versions of 
the points with low $M_S$ and illustrate how the bounds vary with the scalar mass $M_S$ for fixed couplings. Note that, 
for large $M_S$, the spread of the points around the line becomes bigger, which is expected from the LFV/LNV bounds generally 
becoming weaker for large $M_S$. As visible from Figure~\ref{fig:Scenarios}, $\mu^-$--$e^-$ conversion bounds push from top to bottom. 
We have collected several bounds from current and future experiments~\cite{Cui:2009zz,Dohmen:1993mp,Honecker:1996zf,Kutschke:2011ux, Barlow:2011zza}. The 
different scenarios can be constrained depending on the exact values of the model parameters. For example, the \textcolor{blue}{\bf blue} line is easier 
to constrain than the \textcolor{red}{\bf red}/\textcolor{purple}{\bf purple} lines. The reason is that the coefficients $\Pi_{e,\mu,\tau}$, while being 
sufficiently different to spoil cancellations between the three contributions to the total amplitude, are nevertheless all of the same order. Thus, the 
benchmark lines with the largest value of $|f_{ee}^* f_{e\mu}|$, $|f_{e\mu}^* f_{\mu\mu}|$ or $|f_{e\tau}^* f_{\tau\mu}|$ will be easiest to constrain.

\begin{widetext}
\onecolumngrid
\begin{figure}[h!]
\centering
\begin{tabular}{lcr}
\includegraphics[width=6cm]{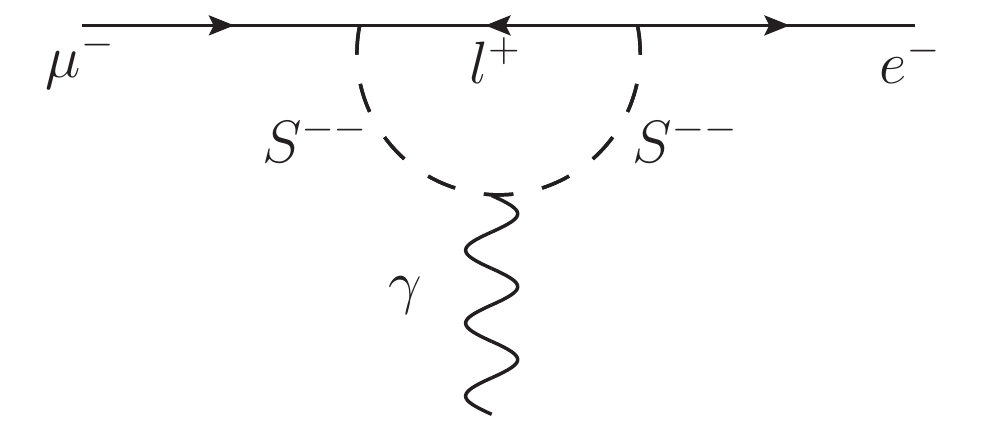} & \includegraphics[width=6cm]{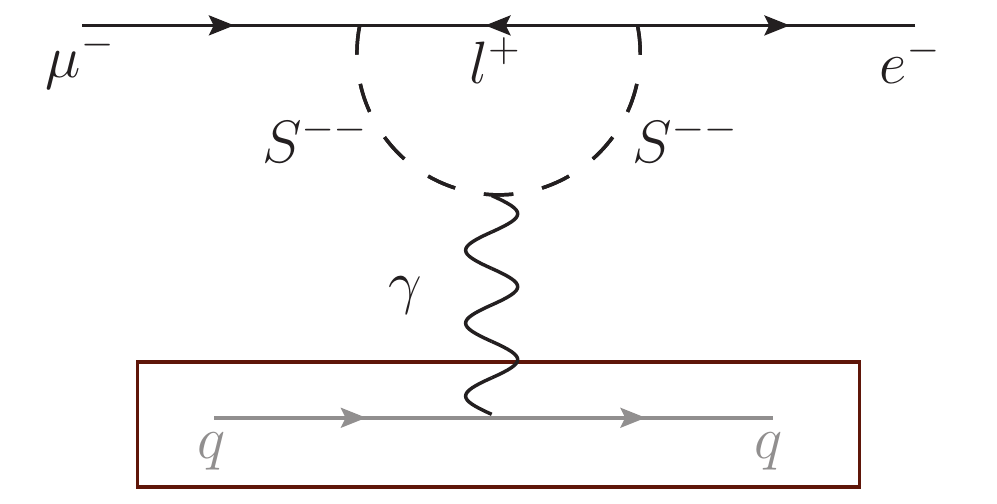} & 
\includegraphics[width=6cm]{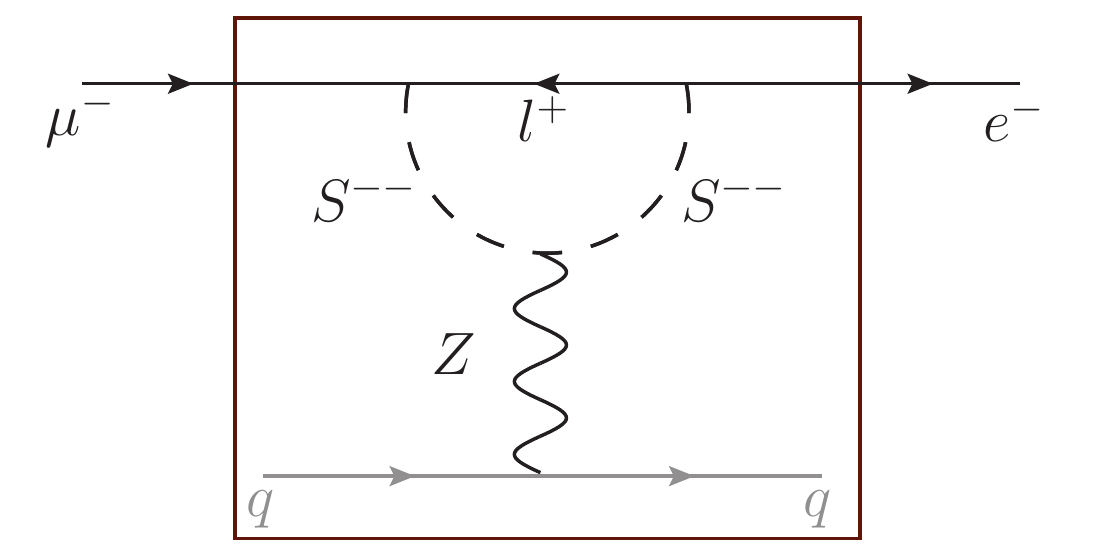}
\end{tabular}
\caption{\label{fig:LFV}Example diagrams for $\mu \to e \gamma$ (left) and $\mu$-$e$ conversion long-range (middle) and short-range (right) 
contributions. The parts inside the large rectangular frames happen inside the nucleus, which necessitates the distinction of long- and short-range. 
}
\end{figure}
\end{widetext}
\twocolumngrid

\vspace{-2mm}

\subsection{Energy Frontier: LHC Searches}
\vspace{-2mm}

\begin{figure}[h!]
\centering
\begin{minipage}{.28\textwidth}
\includegraphics[width=\textwidth]{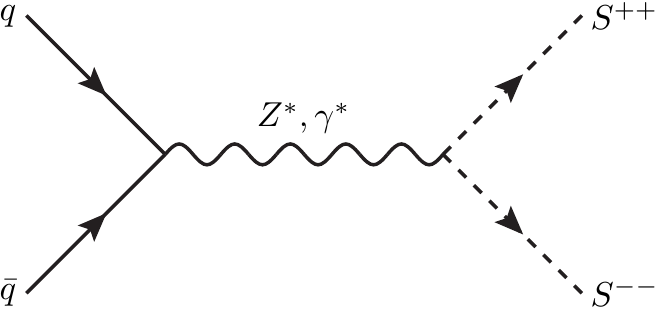}
\end{minipage}

\vspace{5mm}

\begin{minipage}{.22\textwidth}
\includegraphics[width=\textwidth]{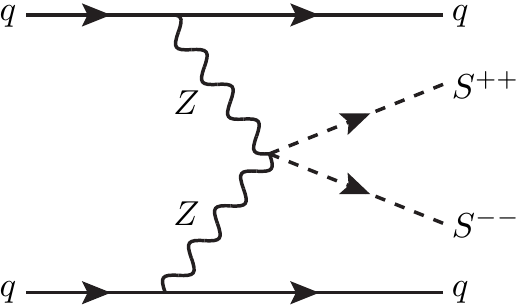}\\
\end{minipage}
\hspace{3mm}
\begin{minipage}{.22\textwidth}
\includegraphics[width=\textwidth]{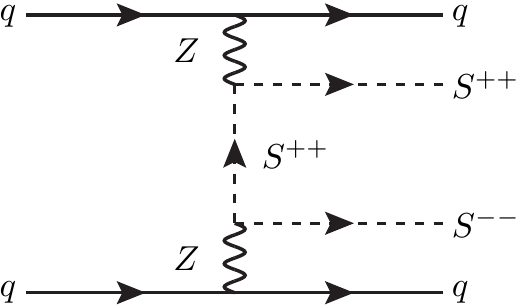}\\
\end{minipage}
\caption{\label{fig:DYVBFFeynman} $S^{++}$ production channels at the LHC: Drell-Yan ({\it Top}); Vector Boson Fusion ({\it Bottom}).}
\end{figure}

Direct searches at the LHC provide a powerful probe of the existence of $S^{++}$ and its antiparticle $S^{--}$, highly complementary 
to $\mu$-$e$ conversion and further LFV processes. At the LHC, the dominant production mode of $S^{++}$ is pair-production 
through the Drell-Yan (DY) process, via a $Z$-boson/photon in the s-channel, as depicted in Figure~\ref{fig:DYVBFFeynman} ({\it Top}). 
Other channels, like pair-production through vector boson fusion (VBF) (see {\it e.g.}~Figure~\ref{fig:DYVBFFeynman} ({\it Bottom}) are largely subdominant, and 
we will not consider them in the present analysis. We however stress that upon discovery of $S^{++}$, these channels could yield valuable 
information on the underlying theory.~Moreover, a potential Higgs portal interaction $\lambda_S |H|^2|S|^2$, as well as the linear interaction (\ref{SppEffLNVO9}) would 
yield new, model-dependent avenues for probing the existence of $S^{++}$ (the interplay between these and the dominant DY production will be explored 
elsewhere \cite{Proj:LHC}).
We nevertheless stress that the effect of the Higgs portal interaction $\lambda_S |H|^2|S|^2$ on the $h \to \gamma\gamma$ decay of the 
125 GeV Higgs is too small to be probed at the LHC for $\lambda_S \lesssim 1$ and/or $m_S \gtrsim 300$ GeV~\cite{Proj:LHC} 

We now analyze the LHC sensitivity to DY production and same-sign di-lepton decays of $S^{++}$ for the three possible coupling patterns 
described above: \textcolor{red}{\bf red}, \textcolor{purple}{\bf purple}, and \textcolor{blue}{\bf blue}. We first concentrate on the 
LHC 7 TeV experimental searches for doubly-charged scalars by ATLAS/CMS~\cite{Chatrchyan:2012ya,ATLAS:2012hi}. Using the 7 TeV Next-to-leading-order 
(NLO) DY production cross-section $\sigma$ values and selection efficiencies $\epsilon_{4\ell}$, $\epsilon_{3\ell\tau}$, $\epsilon_{2\ell2\tau}$ (with $\ell = e,\, \mu$, 
and $\tau$ denoting a hadronically decaying $\tau$-lepton) from~\cite{Chatrchyan:2012ya}, all as a function of $M_S$, we can easily derive the bound on $M_S$ 
that~\cite{Chatrchyan:2012ya} yields for the \textcolor{red}{\bf red}, \textcolor{purple}{\bf purple}, and \textcolor{blue}{\bf blue} benchmarks. The 
number of signal events for each benchmark is given by: 
\begin{eqnarray}
\label{signal_DEF}
s & = & \sigma(M_S) \times \mathcal{L} \nonumber \\ 
&\times &\bigg\{ \mathrm{BR}^2_{\ell\ell} \,\epsilon_{4\ell} + 2\, \mathrm{BR}_{\ell\ell}\,\mathrm{BR}_{\ell\tau} 
\left[br_{\tau_{\ell}}\,\epsilon_{4\ell} + br_{\tau_H}\, \epsilon_{3\ell\tau}\right] \nonumber\\
&+& \mathrm{BR}^2_{\ell\tau} \left[ br^2_{\tau_{\ell}}\,\epsilon_{4\ell} + 2\, br_{\tau_H}\, br_{\tau_{\ell}}\, \epsilon_{3\ell\tau} + 
br^2_{\tau_H}\, \epsilon_{2\ell2\tau} \right] \bigg\},
\end{eqnarray}
where $\mathrm{BR}_{\ell\ell}$ and $\mathrm{BR}_{\ell\tau}$ are, respectively, the $S^{++}$ branching fractions into two light 
leptons ($\ell = e,\,\mu$) and into a light lepton and a $\tau$-lepton, which can be directly obtained as ratios of squared couplings from Tab.~\ref{Tab:Scenarios}. 
The hadronic and leptonic branching fractions of a $\tau$-lepton 
are respectively given by $br_{\tau_H} \simeq 0.65$, $br_{\tau_{\ell}} \simeq 0.35$.

The $M_S$ limit is then obtained using the $\mathrm{CL}_s$ procedure~\cite{Read:2000ru,Read:2002hq}, for which we construct the 
likelihood ratio test-statistics $Q$, corresponding to the ratio of likelihoods for the signal+background ($s+b$) and background 
only ($b$) hypotheses for the observed number of events in each experimental bin $n_i$, and then compute the exclusion confidence 
level of the signal under the assumption of the $b$ hypothesis by comparing the $p$-values of the Poissonian distributions of the $s+b$ and $b$ hypotheses:
\begin{equation}
\mathrm{CL}_s=\frac{1-p(s+b)}{1-p(b)}\quad, \quad Q = e^{-s} \prod_{i=1}^{N_{\mathrm{bins}}} \left(1+\frac{s_i}{b_i}\right)^{n_i},
\end{equation}

For the LHC 13 TeV expected limits, we have computed the Leading Order DY pair-production cross sections with 
MadGraph5\_aMC@NLO \cite{Alwall:2014hca} and obtained the rescaled NLO cross-sections via an average, $M_S$-independent, 
$\kappa$-factor of $1.25$~\cite{Muhlleitner:2003me}. We then consider both an ideal scenario, 
with 100\% signal selection efficiencies, and a conservative one in which we extrapolate to LHC $13$ TeV the values for 
the efficiencies $\epsilon_{4\ell},\epsilon_{3\ell\tau},\epsilon_{2\ell2\tau}$ at $7$~TeV (which are expected to improve 
from LHC Run~1 to Run~2). In both cases, we use (\ref{signal_DEF}) and the $\mathrm{CL}_s$ method under the hypothesis of no 
background events (and therefore no observed events). To obtain a 95\%~CL exclusion under these hypotheses, the number of 
signal events must be larger than 3. The results for the $M_S$ bounds at both $7$~TeV and $13$~TeV are shown in Tab.~\ref{tab:LHCbounds}, 
and then included in Figure~\ref{fig:Scenarios}, which clearly shows the complementarity between LFV and LHC searches for the three benchmarks.

\begin{table}
\hspace{-4mm}
\begin{tabular}{cc|c|ccc}
\hline
\begin{tabular}{c}Energy\\(TeV)\end{tabular}&
\begin{tabular}{c}$\mathcal{L}$\\(fb)\end{tabular}&
\begin{tabular}{c}Ideal\\Bound\end{tabular}&
\begin{tabular}{c}\textcolor{red}{\bf red}\\$\mathrm{BR}_{\ell\ell}=1$\end{tabular} & 
\begin{tabular}{c}\textcolor{purple}{\bf purple}\\$\mathrm{BR}_{\ell\tau}=0.98$\end{tabular} & 
\begin{tabular}{c}\textcolor{blue}{\bf blue}\\$\mathrm{BR}_{\ell\ell}=0.99$\end{tabular} \\
\hline
7 & 4.9 & 423 & 364 & 293 & 363 \\
\hline
\multirow{2}{*}{13} & 100 & 900 & 781 & 664 & 780 \\
& 300 & 1102 & 977 & 811 & 976 \\
\hline
\end{tabular}
\caption{\label{tab:LHCbounds}LHC $M_S$ bounds (in GeV) for the different benchmarks. The ideal bound assumes 100\% efficiencies 
for all channels ($\epsilon_{4\ell},\epsilon_{3\ell\tau},\epsilon_{2\ell2\tau} = 1$).}
\end{table}


\begin{widetext}
\onecolumngrid
\begin{figure}[h!]
\begin{tabular}{lcr}
\hspace{-0.5cm}
\includegraphics[width=6.cm]{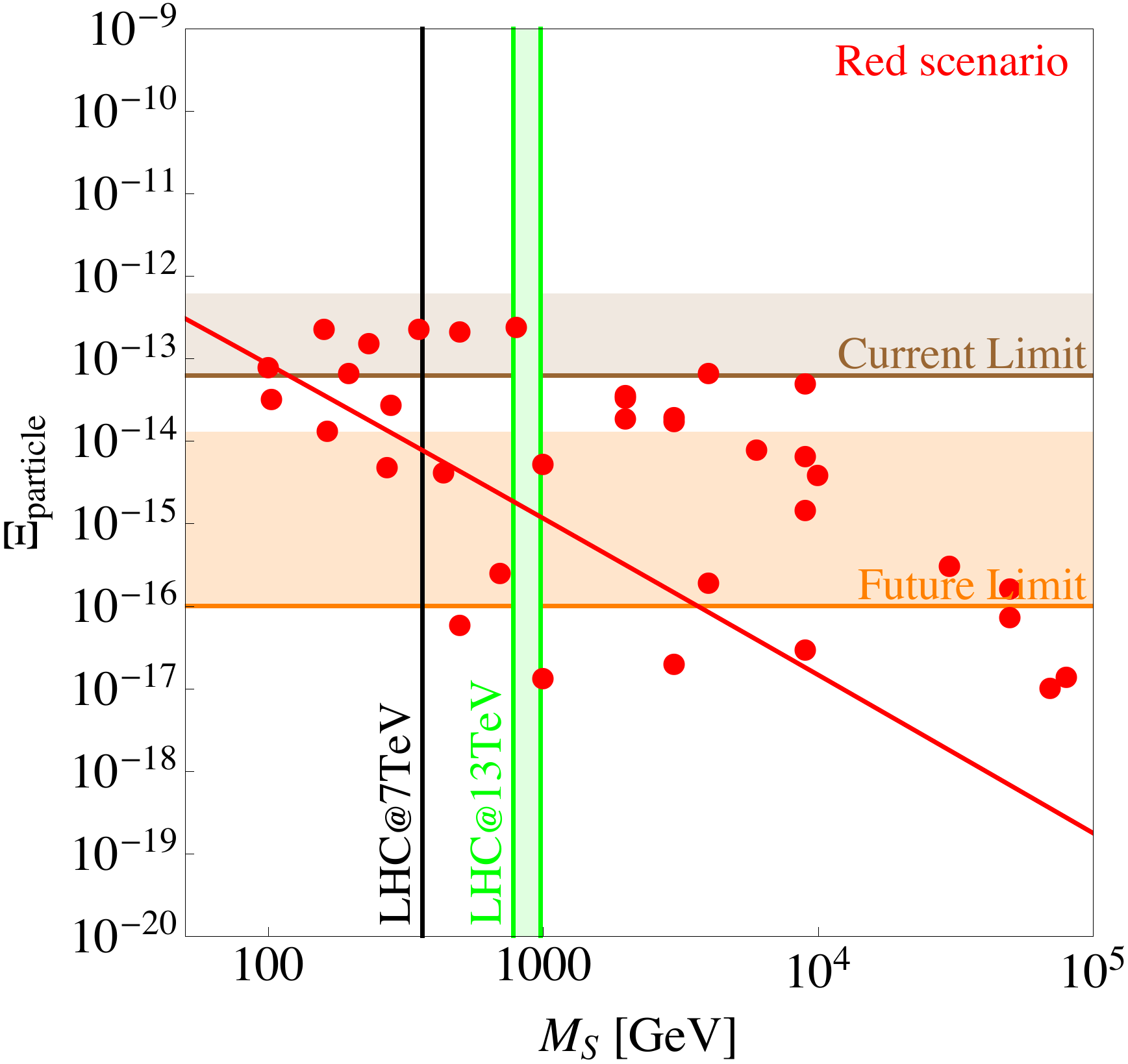}
&
\includegraphics[width=6.cm]{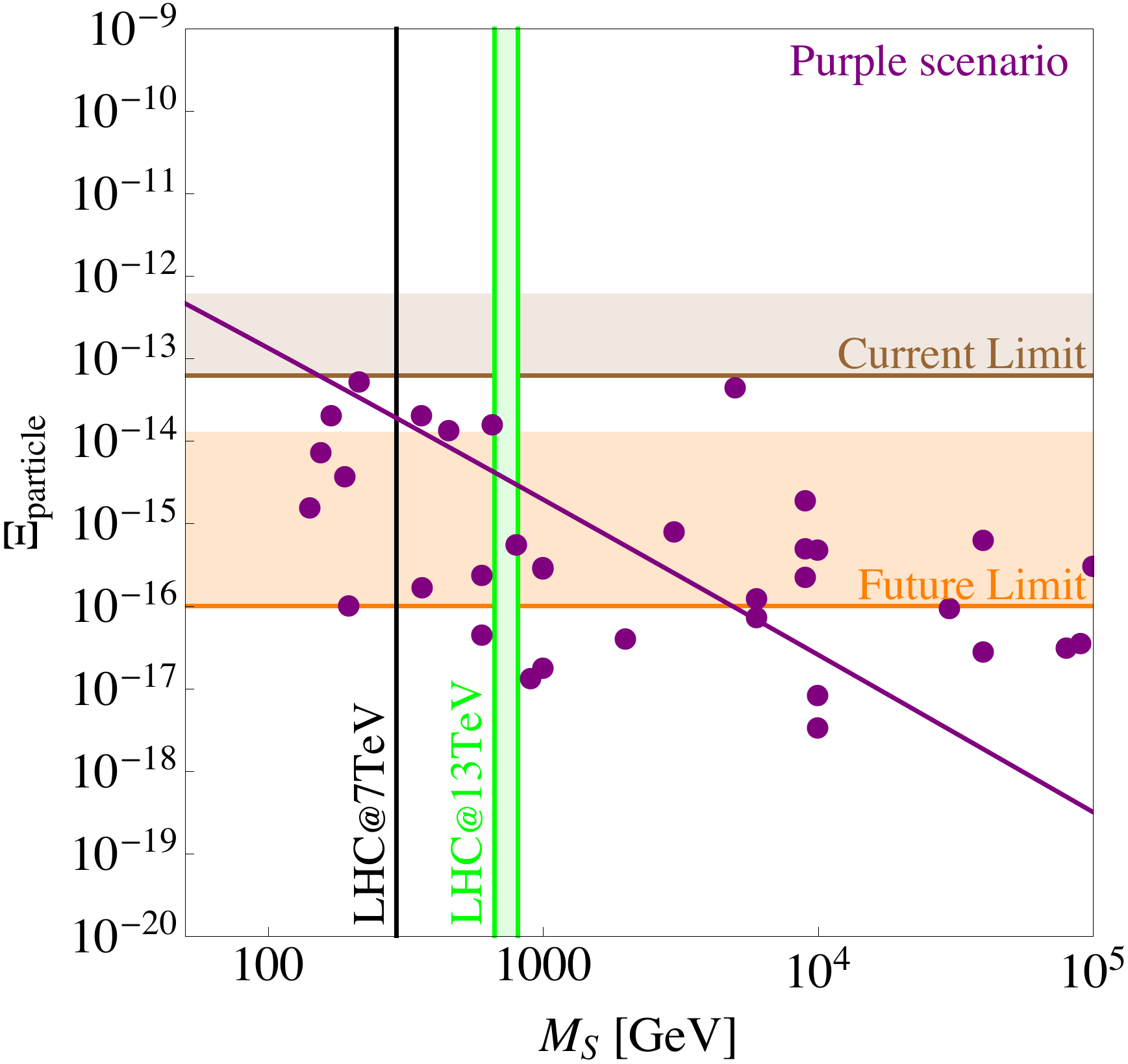}
&
\includegraphics[width=6.cm]{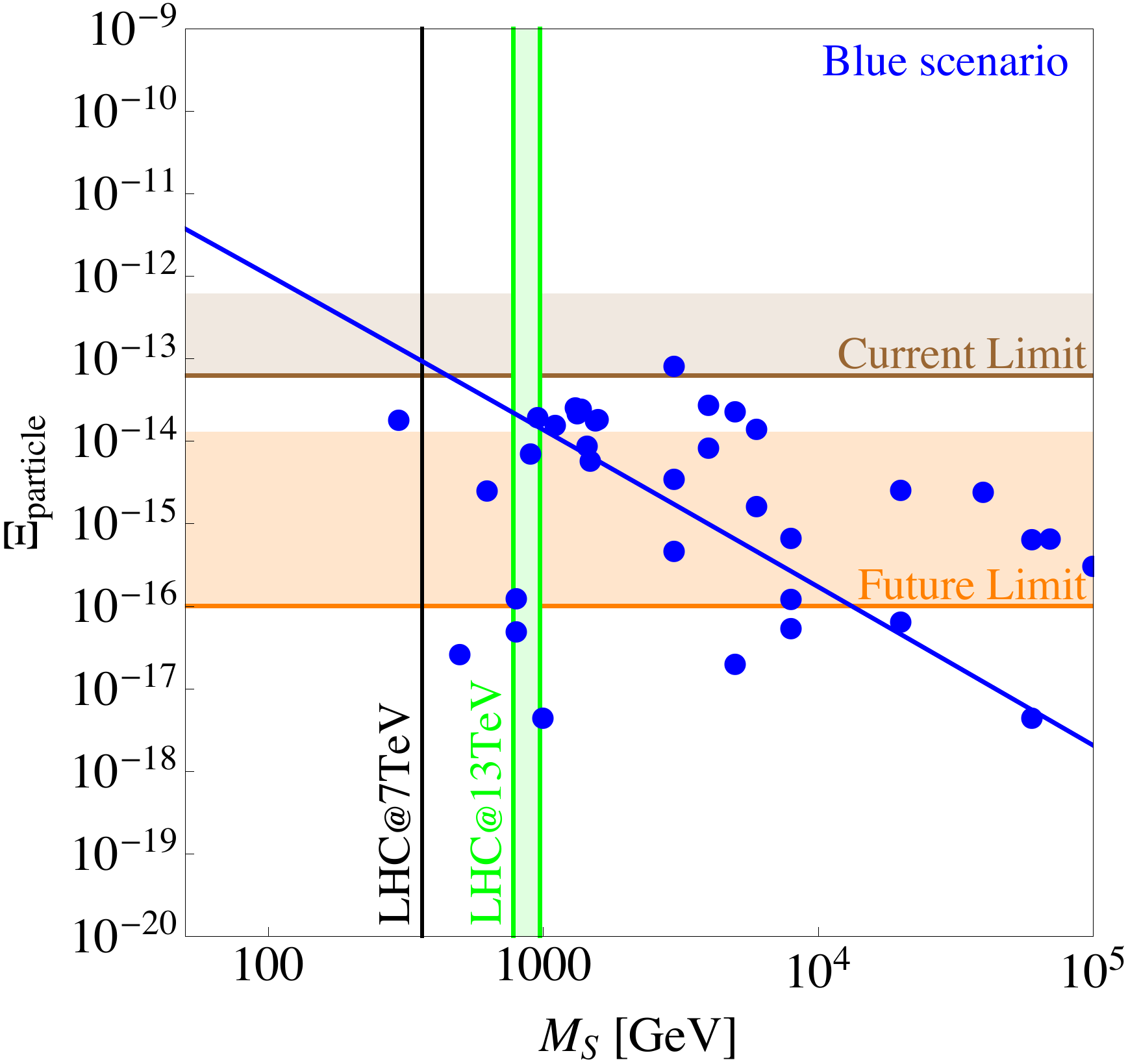}
\end{tabular}
\caption{\label{fig:Scenarios}Particle physics parts $\Xi_{\rm particle}$ as functions of the scalar mass $M_S$ for the 
red/purple/blue scenarios. Different limits from $\mu^-$--$e^-$ conversion and from LHC searches are indicated, see text for details.}
\end{figure}
\end{widetext}

\section{Conclusion}

\vspace{-2mm}

We have shown how the intensity and the energy frontiers provide complementary constraints within a minimal model of 
neutrino mass and mixing involving just one new particle beyond the Standard Model, namely a doubly charged scalar $S^{++}$ and 
its antiparticle $S^{--}$. Focussing on the complementarity between LHC searches and low energy probes such as lepton flavor violation, 
the results are summarised in Figure~\ref{fig:Scenarios}. The complementary nature of these approaches is very clear from this figure, with 
the LHC able to exclude scalar masses approaching 1 TeV, while $\mu^-$--$e^-$ conversion holds the promise of orders of magnitude improvement 
in constraining the particle physics amplitude in~\eqref{eq:muecon_particle-part}, albeit with rather large uncertainties due to nuclear physics.

From a general perspective, the considered framework provides a minimal and clear example of complementarity between two of the most 
important experimental particle physics strategies presently being pursued, towards uncovering the physics beyond the Standard Model, 
which must necessarily be present to account for neutrino mass and mixing.

{\bf Note added:} As this paper was being submitted,~\cite{Chakrabortty:2015zpm} appeared, which also discusses the importance of 
the complementarity of high- and low-energy data for the case of doubly charged scalars. While that paper also discusses several LFV aspects, 
its main focus is the muon magnetic  dipole moment $(g-2)$, while our paper focuses strongly on the complementarity between LHC direct searches and $\mu$-$e$ conversion.

\vspace{4mm}

\begin{acknowledgments}
\paragraph{Acknowledgements}
We are indebted to Richard Ruiz for very useful comments on a first version of this work.
SFK acknowledges partial support from the STFC Consolidated ST/J000396/1 grant. AM acknowledges warm hospitality by the 
University of Southampton, where part of this work has been performed, as well as partial support by the Micron Technology Foundation, Inc. SFK and AM furthermore acknowledge partial support by the 
European Union FP7 ITN-INVISIBLES (Marie Curie Actions, PITN-GA-2011-289442). JMN is supported by the People Programme (Marie curie Actions) of the 
European Union Seventh Framework Programme (FP7/2007-2013) under REA grant agreement PIEF-GA-2013-625809.
\end{acknowledgments}


\bibliographystyle{apsrev}
\bibliography{Complementarity_PRL}


\end{document}